\documentclass[10pt]{article}
\usepackage{multicol}
\usepackage{color}
\usepackage{booktabs}
\usepackage{tabu}
\usepackage{float}

\usepackage{ntheorem}
\usepackage[T1]{fontenc}

\usepackage{titling}
\usepackage{enumerate}
\usepackage[centertags]{amsmath}
\usepackage{graphicx}  
\usepackage{newlfont}
\usepackage{dsfont}
\usepackage{amssymb,amsmath,amsfonts}
\usepackage{epstopdf}
\usepackage[bf,normalsize,center]{subfigure}
\usepackage{multirow}
\usepackage{titlesec}
\usepackage{empheq}
\usepackage{cuted}
\usepackage{mathtools}
\usepackage{caption}
\usepackage{setspace}

\makeatletter
\renewcommand*\env@matrix[1][*\c@MaxMatrixCols c]{%
  \hskip -\arraycolsep
  \let\@ifnextchar\new@ifnextchar
  \array{#1}}
\makeatother

\DeclareMathOperator*{\argmin}{arg\,min}

\usepackage{relsize}

\usepackage{geometry}
\usepackage[dvipsnames]{xcolor}
\usepackage{empheq}

\newcommand{\Real}{{\mathds R}} 
\newcommand{\Nat}{{\mathds N}} 

\newtheorem{definition}{Definition}{}
\newtheorem{corollary}{Corollary}{}
\newtheorem{proposition}{Proposition}{}
\newtheorem{problem}{Problem}{}
\newtheorem{theorem}{Theorem}{}
\newtheorem{remark}{Remark}{}
\newtheorem{lemma}{Lemma}{}
{}

\usepackage{graphicx}
\newcommand{\indep}{\rotatebox[origin=c]{90}{$\models$}}

\usepackage{soul}

\titlespacing*{\section}
{0pt}{1.0ex plus 1ex minus .2ex}{1.0ex plus .2ex}
\titlespacing*{\subsection}
{0pt}{1.0ex plus 1ex minus .2ex}{1.0ex plus .2ex}

\makeatletter
\g@addto@macro\normalsize{%
  \setlength\abovedisplayskip{3pt}
  \setlength\belowdisplayskip{3pt}
}
\makeatother

\DeclareMathOperator*{\supess}{ess\,sup}

\begin{document}

\setul{}{1pt}
\noindent
\Large
\textbf{Information\hspace{1mm}-Theoretic Privacy through Chaos Synchronization  and\\[1.5mm] Optimal Additive Noise}\\[2mm]\large
Carlos Murguia$^{1,a}$, Iman Shames$^{1,b}$, Farhad Farokhi$^{1,2,c}$, and Dragan Ne\v{s}i\'{c}$^{1,d}$\\[3mm]\normalsize
$^1$Department of Electrical and Electronic Engineering, University of Melbourne, Australia\\[2mm]
$^2$The Commonwealth Scientific and Industrial Research Organisation (CSIRO), Data61, Australia\\[2mm]
\ul{$^a$carlos.murguia@unimelb.edu.au}; \ul{$^b$iman.shames@unimelb.edu.au}; \ul{$^c$farhad.farokhi@unimelb.edu.au};\\ \ul{$^d$dnesic@unimelb.edu.au}

\normalsize


\section{Abstract}

We study the problem of maximizing privacy of data sets by adding random vectors generated via synchronized chaotic oscillators. In particular, we consider the setup where information about data sets, queries, is sent through public (unsecured) communication channels to a remote station. To hide private features (specific entries) within the data set, we corrupt the response to queries by adding random vectors. We send the distorted query (the sum of the requested query and the random vector) through the public channel. The distribution of the additive random vector is designed to minimize the mutual information (our privacy metric) between private entries of the data set and the distorted query. We cast the synthesis of this distribution as a convex program in the probabilities of the additive random vector. Once we have the optimal distribution, we propose an algorithm to generate pseudorandom realizations from this distribution using trajectories of a chaotic oscillator. At the other end of the channel, we have a second chaotic oscillator, which we use to generate realizations from the same distribution. Note that if we obtain the same realizations on both sides of the channel, we can simply subtract the realization from the distorted query to recover the requested query. To generate equal realizations, we need the two chaotic oscillators to be synchronized, i.e., we need them to generate exactly the same trajectories on both sides of the channel synchronously in time. We force the two chaotic oscillators into \emph{exponential synchronization} using a driving signal. Exponential synchronization implies that trajectories of the oscillators converge to each other exponentially fast for all admissible initial conditions and are perfectly synchronized in the limit only. Thus, in finite time, there is always a ``small'' difference between their trajectories. To implement our algorithm, we assume (as it is often done in related work) that systems have been operating for sufficiently long time so that this small difference is negligible and oscillators are practically synchronized. We quantify the worst-case distortion induced by assuming perfect synchronization, and show that this distortion vanishes exponentially fast. Simulations are presented to illustrate our results.\\[2mm]
\textbf{Keywords: Privacy; Data Sets, Queries, Mutual Information, Chaos.}

\section{Introduction}

In a hyperconnected world, scientific and technological advances have led to an overwhelming amount of user data being collected and processed by hundreds of companies over public networks. Companies mine this data to provide targeted advertising and personalized services. However, these new technologies have also led to an alarming widespread loss of privacy in society. Depending on adversary's resources, opponents may infer private user information from public data available on the internet and unsecured/public servers. A motivating example of privacy loss is the potential use of data from smart electrical meters by criminals, advertising agencies, and governments, for monitoring the presence and activities of occupants \cite{Poor1,Poor2}. Other examples are privacy loss caused by information sharing in distributed control systems and cloud computing \cite{Huang:2014:CDP:2566468.2566474}; the use of travel data for traffic estimation in intelligent transportation systems \cite{Gruteser}; and data collection and sharing by the Internet-of-Things (IoT) \cite{WEBER201023}, which is, most of the time, done without the user's informed consent. These privacy concerns show that there is an acute need for privacy preserving mechanisms capable of handling the new privacy challenges induced by an interconnected world.

In this manuscript, we consider the problem of hiding private information $X$ of users (modeled as discrete random vectors) within datasets when publicly sharing requested queries $Y(X)$ from the same source. In particular, the aim of our privacy scheme is to respond to queries with distorted queries of the form $Z=Y(X)+V$ such that, when releasing $Z$, the private $X$ is ``hidden''. Realizations of the vector $Z$ are transmitted over a public (unsecured) communication channel to a remote station. Then, if we do not distort $Y(X)$ before transmission, information about $X$ is directly accessible through the public channel. The first problem that we address is the design of the probability distribution of $V$ to maximize privacy, i.e., the distribution of $V$ must be constructed so that $Z = Y(X) + V$ carries as little information about $X$ as possible. Here, we follow an \emph{information-theoretic approach} to privacy. We use the \emph{mutual information} between private information $X$ and distorted queries $Y(X) + V$, $I[X;Y(X) + V]$, as \emph{privacy metric}. The design of the discrete additive vector is casted as an optimization problem where we minimize $I[X;Y(X) + V]$ using the probability mass function of $V$, $p_V(v)$, as optimization variables. That is, the optimal distribution, $p^*_V(v)$, is given by $p^*_V(v) := \argmin_{p_V(v)} I[X;Y(X) + V]$, where $p_V(v)$ is taken over a class of probability mass functions. Contrary to related work \cite{Topcu}-\nocite{SORIA}\nocite{Geng}\nocite{Fawaz}\nocite{Carlos_Iman1}\cite{Dullerud}, we do not consider any sort of privacy-distortion trade-off in our formulation. We actually aim at making $I[X;Y(X) + V]$ as small as possible regardless of the distortion between $Y(X)$ and $Y(X) + V$ induced by $V$. Distortion is not an issue because we seek to generate exactly the same realization of $V$ at the remote station; then, we could recover the query by simply subtracting this realization from the one of $Z=Y(X) + V$. In order to accomplish this, we propose an algorithm to generate pseudorandom realizations from $p^*_V(v)$ at both sides of the channel using trajectories of two synchronized chaotic oscillators.

There are a number of requirements that the oscillators must satisfy for our algorithm to work: 1) trajectories of the oscillators must be \emph{bounded} and \emph{chaotic}; 2) they must be \emph{synchronized}, i.e., we need them to generate exactly the same trajectories on both sides of the channel synchronously in time; and 3) the synchronous solution, regarded as a random process, must be \emph{stationary}. Before giving the algorithm, we provide general guidelines for selecting the dynamics of the oscillators so that all the aforementioned requirements are satisfied. In particular, we use a range of well-known results in the literature to provide a synthesis procedure that allows to choose suitable oscillators. For boundedness, we use the notion of \emph{Input-to-State-Stability} (ISS); for chaos, we employ standard \emph{largest Lyapunov exponent methods} \cite{wiggins} and the (0-1) \emph{test} \cite{Melbourne}; for synchronization, we introduce the notion of \emph{convergent systems} \cite{Pav}; and for stationarity, we use \emph{hyperbolicity} of the chaotic trajectories \cite{Anishchenko}.

To generate equal realizations, our algorithm needs trajectories of the two chaotic oscillators (one at each side of the channel) to be synchronized. We force the oscillators into \emph{exponential synchronization} using a driving signal. Exponential synchronization implies that trajectories of the oscillators converge to each other exponentially for all admissible initial conditions and are perfectly synchronized in the limit only. Therefore, in finite time, there is always a ``small'' difference between their trajectories. However, because oscillators synchronize exponentially fast, and it is often possible in practice to select initial conditions from a known compact set (known to both sides of the channel), it is safe to assume that the interconnected systems have been operating for sufficiently large time such that oscillators are \emph{practically} synchronized, i.e., the synchronization error is so small that trajectories can be assumed to be equal. This is a standard assumption that is made in most, if not all, of the existing work on chaotic encryption based on synchronization \cite{ChuaChao1}-\nocite{GRZYBOWSKI20092793}\nocite{LU2002365}\nocite{ALVAREZ2005775}\cite{ChuaChao2}. Here, we give sufficient conditions for exponential synchronization to occur, provide tools for selecting the oscillators such that these conditions are satisfied, and assume that, after transients have settled down, trajectories are perfectly synchronized to some chaotic trajectory, say $\phi(t) \in \Real^{n_\zeta}$, $\zeta \in \Nat$. If $n_\zeta>1$, our algorithm uses any entry $\phi^s(t) \in \mathcal{S} \subset \Real$ of $\phi(t)$ to generate realizations from $p^*_V(v)$, where $\mathcal{S}$ denotes some compact set that characterizes the support of $\phi^s(t)$. Because oscillators are selected such that $\phi(t)$, regarded as a random process, is stationary, samples from $\phi^s(t)$ follow a stationary probability density function. We obtain this density through Monte Carlo simulations \cite{Robert:2005:MCS:1051451} and  divide its support $\mathcal{S}$ into a finite set of cells $C = \{c^1,\ldots,c^M\}$ such that the probability that $\phi^s(t)$ lies in these cells equals the optimal probability distribution $p_V^*(v)$. That is, we generate pseudorandom realizations from $p_V^*(v)$ by properly selecting $C$ and evaluating if $\phi^s(t)$ lies in $C$ at the sampling instants.

The use of additive noise to preserve privacy is common practice. There are mainly two classes of privacy metrics considered in the literature; namely, differential privacy \cite{Jerome1}-\cite{Dwork} and information-theoretic metrics, e.g., mutual information, conditional entropy, Kullback-Leibler divergence, and Fisher information \cite{Farokhi1}-\nocite{Farokhi2}\nocite{FAROKHI3}\nocite{Fawaz2}\cite{Poor}. In differential privacy, because it provides certain privacy guarantees, Laplace noise is usually used \cite{Dwork2}. However, when maximal privacy is desired, Laplace noise is generally not the optimal solution. This raises the fundamental question: what is the noise distribution achieving maximal privacy? This question has many possible answers depending on the particular privacy metric being considered and the system configuration, see, e.g., \cite{Topcu}-\nocite{SORIA}\cite{Geng},\cite{Dullerud}, for differential privacy based results, and \cite{Farokhi1}-\nocite{Farokhi2}\nocite{FAROKHI3}\nocite{Fawaz2}\cite{Poor}, for information theoretic results. In general, if the data to be kept private follows continuous distributions, the problem of finding the optimal additive noise to maximize privacy is hard to solve. If a close-form solution for the distribution is desired, the problem amounts to solving a set of nonlinear partial differential equations which, in general, might not have a solution, and even if they do have a solution, it is hard to find \cite{Farokhi1}. This problem has been addressed by imposing some particular structure on the considered distributions or assuming the data to be kept private is deterministic \cite{Farokhi1},\cite{SORIA},\cite{Geng}. The authors in \cite{SORIA},\cite{Geng} consider deterministic input data sets and treat optimal distributions as distributions that concentrate probability around zero as much as possible while ensuring differential privacy. Under this framework, they obtain a family of piecewise constant probability density functions that achieve minimal distortion for a given level of privacy. In \cite{Farokhi1}, the authors consider the problem of preserving the privacy of deterministic databases using additive continuous noise with constrained support. They use the Fisher information and the Cramer-Rao bound to construct a privacy metric between deterministic data and the one with the additive noise, and find the probability density function that minimizes it. Moreover, they prove that, in the unconstrained support case, the optimal noise distribution minimizing the Fisher information is Gaussian. This observation has been also made in \cite{Cedric} when using mutual information as a measure of privacy. We remark that most of the aforementioned papers consider privacy-distortion trade-offs when designing their distorting mechanisms. We do not consider this trade-off here because, at the end of the channel, we remove the distortion that we induce using our synchronization based formulation.

Existing work on chaotic encryption based on synchronization \cite{ChuaChao1}-\nocite{GRZYBOWSKI20092793}\nocite{LU2002365}\nocite{ALVAREZ2005775}\cite{ChuaChao2} directly uses the states of the chaotic oscillators to mask private information. That is, standard algorithms do not use chaotic trajectories to generate pseudorandom realization from probability distributions (as we do here); instead, they simply add the value of the sampled chaotic trajectory (or functions of it) to private messages. Although the latter succeeds in masking messages, it does not give any privacy guarantees (neither information-theoretic nor in a differential privacy sense) on the private information, and it is not optimal in any sense.  Hence, the contributions of our scheme with respect to existing work on chaotic encryption \cite{ChuaChao1}-\nocite{GRZYBOWSKI20092793}\nocite{LU2002365}\nocite{ALVAREZ2005775}\cite{ChuaChao2} are the treatment of fully stochastic datasets, the information-theoretic privacy guarantees that our framework provides, and the optimal performance of the designed distorting additive vector (optimal in the sense of minimizing the mutual information $I[X;Y(X) + V]$). The work here is inspired by the experimental results presented in \cite{Lars}, where the authors propose a framework similar to ours for deterministic data using a electronic circuit implementation of the Mackey-Glass chaotic oscillator \cite{Mackey287}. The contribution of our work with respect to \cite{Lars} is threefold: 1) we consider fully stochastic data, which makes the privacy scheme fundamentally very different; 2) we provide a general formulation that encompasses a large class of chaotic systems, not only the electronic circuit implementation of the Mackey-Glass oscillator; and 3) we generate realizations from optimal distorting distributions, in \cite{Lars}, they consider uniform distributions only which is not optimal for stochastic data.\vspace{1mm}

Next, we summarize the main contributions of the chapter.\\[1mm]
\textbf{Contributions:}\\[1mm]
1) We provide a general information-theoretic privacy framework based on optimal additive distorting random vectors and synchronization of chaotic oscillators; 2) We prove that the synthesis of the probability mass function $p_V(v)$ of the distorting random vector $V$ can be posed as a convex program in $p_V(v)$ over a class of probability mass functions; 3) We provide an algorithm to generate pseudorandom realizations from this distribution using trajectories of chaotic oscillators; 4) Using off-the-shelf results in the literature, we provide a synthesis procedure for selecting the dynamics of the oscillators so that our algorithm is guaranteed to work.

The remainder of the paper is organized as follows. In Section 3, we present some preliminaries results needed for the subsequent sections. We introduce the notion of convergent systems and the concept of mutual information. The general formulation and the specific problems to be addressed are given in Section 4. In Section 5, we pose the synthesis of the probability distribution of the optimal distorting vector. General guidelines for selecting the chaotic oscillators are given in Section 6. The algorithm for generating pseudorandom realizations from the optimal distribution is presented in Section 7. Simulation results are given in Section 8 and concluding remarks in Section 9.

\section{Notation and Preliminaries}\label{Prelim}

The symbol $\Real$ stands for the real numbers, $\Real_{>0}$($\Real_{\geq 0}$) denotes the set of positive (non-negative) real numbers. The symbol $\Nat$ stands for the set of natural numbers. The Euclidian norm in $\Real^n$ is denoted simply as $|\cdot|$, $|x|^2=x^\top x$, where $^\top$ defines transposition. For a given measurable function $u(t)$, $t \in \Real_{\geq 0}$, we denote its $\mathcal{L}_{\infty}$ norm as $||u||_{\infty} := \supess_{t \geq 0}|u(t)|$, where $\supess$ denotes essential supremum. Matrices composed of only ones and only zeros of dimension $n \times m$ are denoted by $\mathbf{1}_{n \times m}$ and $\mathbf{0}_{n \times m}$, respectively, or simply $\mathbf{1}$ and $\mathbf{0}$ when their dimensions are clear. For square matrices $A\in \Real^{n \times n}$, $\rho[A]$ denotes the spectral radius of $A$. A continuous function $\gamma:[0,a) \rightarrow [0,\infty)$ is said to belong to class $\mathcal{K}$ if it strictly increasing and $\gamma(0)=0$. Similarly, a continuous function $\beta:[0,a) \times [0,\infty) \rightarrow [0,\infty)$ belongs to class $\mathcal{KL}$ if, for fixed $s$, $\beta(r,s)$ belongs to class $\mathcal{K}$ with respect to $r$ and, for fixed $r$, $\beta(r,s)$ is decreasing with respect to $s$ and $\lim_{s \rightarrow \infty} \beta(r,s) = 0$. Consider a discrete random vector $X$ with alphabet $\mathcal{X}=\{x_1,\ldots,x_N\}$, $x_i \in \Real^m$, $m \in \Nat$, $i\in \{1,\ldots,N\}$, and probability mass function $p_X(x) = \text{Pr}[X=x]$, $x \in \mathcal{X}$, where $\text{Pr}[B]$ denotes probability of event $B$. Similarly, for two random vectors $X$ and $Y$, taking values in the alphabets $\mathcal{X}$ and $\mathcal{Y}$, respectively, their joint probability mass function is denoted by $p_{X,Y}(x,y)$, the marginal distribution of $X$ is given by $p_X(x) = \sum_{y \in \mathcal{Y}}p_{X,Y}(x,y)$, and the conditional distribution of $X$ given $Y$ as $p_{Y|X}(y|x)=p_{X,Y}(x,y)/p_X(x)$. Analogously, for a continuous random vector $Y$, we denote their (multivariate) probability density function as $f_Y(y)$. The notation $X \sim f_X(x)$ ($X \sim p_X(x)$) stands for continuous (discrete) random vectors $X$ following the probability density (mass) function $f_X(x)$ ($p_X(x)$). We denote by "Simplex" the probability simplex defined by $\sum_{x \in \mathcal{X}}p_X(x) = 1$, $p_X(x) \geq 0$ for all $x \in \mathcal{X}$. The notation $E[a]$ denotes the expected value of the random vector $a$. We denote independence between two random vectors, $X$ and $Y$, as $X \indep Y$.\\

\subsection{Mutual Information}

\begin{definition}\label{mutual_info}
Consider two random vectors, $X$ and $Y$, with joint probability mass function $p_{X,Y}(x,y)$ and marginal probability mass functions, $p_X(x)$ and $p_Y(y)$, respectively. Their mutual information $I[X;Y]$ is defined as the relative entropy between the joint distribution and the product distribution $p_X(x)p_Y(y)$, i.e.,
\[
I[X;Y]:= \sum_{x \in \mathcal{X}}\sum_{y \in \mathcal{Y}}p_{X,Y}(x,y)\log \frac{p_{X,Y}(x,y)}{p_X(x)p_Y(y)}.
\]
\end{definition}

Mutual information $I[X;Y]$ between two jointly distributed vectors, $X$ and $Y$, is a measure of the dependence between $X$ and $Y$.\\

\subsection{Convergent Systems}
Consider the dynamical system:
\begin{equation}
\dot{x}(t) = r(x(t),u(t)),  \label{1}
\end{equation}
with $t \in\Real_{\geq 0}$, state $x \in \Real^n$, input $u \in \mathcal{U} \subseteq \Real^m$, and vector field $r:\Real^n \times \mathcal{U} \rightarrow \Real^n$. The vector field $r(x,u)$ is continuously differentiable in $x$, and $u(t)$ is piecewise continuous in $t$ and takes values in some compact set $\mathcal{U} \subseteq \Real^m$.

\begin{definition}{\emph{\cite{Demi}}.}\label{converg}
System \emph{(\ref{1})} is said to be globally asymptotically convergent if and only if for any bounded input $u(t)$, $t \in \Real$, there is a unique bounded globally asymptotically stable solution $\bar{x}_u(t)$, $t \in \Real$, such that $\lim_{t\rightarrow \infty }\left\vert x(t)-\bar{x}_u(t)\right\vert =0$ for all initial conditions.
\end{definition}
For a \emph{convergent system}, the limit solution is solely determined by the external excitation $u(t)$ and not by the initial conditions. A sufficient condition for convergence obtained by Demidovich \cite{Demi} and later extended in \cite{Pav} is presented in the following proposition.

\begin{proposition}{\emph{\cite{Demi,Pav}}.}\label{converg2}
If there exists a positive definite matrix $P \in \Real^{n \times n}$ such that all the eigenvalues $\lambda_i(Q)$ of the symmetric matrix
\begin{equation}
Q(x,u)=\frac{1}{2} \left( P\left( \frac{\partial r}{\partial x }%
(x,u)\right) +\left( \frac{\partial r}{\partial x }(x
,u)\right) ^{T} P \right),   \label{7P}
\end{equation}
are negative and separated from zero, i.e., there exists a constant $c \in \Real_{>0}$ such that $\lambda_i (Q)\leq -c <0,$ for all  $i \in \{1,...,n \}$, $u \in \mathcal{U}$, and $x \in \Real^n$, then system \emph{(\ref{1})} is globally exponentially convergent; and, for any pair of solutions $x_1(t),x_2(t) \in \Real^n$ of \emph{(\ref{1})}, the following is satisfied:
\[
\frac{d}{dt}\Big(\big( x_{1}(t)-x_{2}(t) \big)^\top P\big( x_{1}(t)-x_{2}(t) \big)\Big)\leq -\alpha \left\vert
 x_{1}(t) - x_{2}(t) \right\vert ^{2}, \hspace{1mm}t \in \Real_{\geq 0},
\]
with constant $\alpha := (c/\lambda_{\max}(P))$ and $\lambda_{\max}(P)$ being the largest eigenvalue of the symmetric matrix $P$.
\end{proposition}

\begin{remark}\label{incrementalISS}
There are other methods to verify that trajectories of system \eqref{1} converge to a limit solution that is independent of the initial conditions and solely determined by the external excitation $u(t)$. For instance, contraction theory \emph{\cite{Slotine}}, Lyapunov function approach to incremental stability \emph{\cite{Angeli}}, the quadratic (\emph{QUAD}) inequality approach (a Lipschitz-like condition) \emph{\cite{QUAD}}, and differential dissipativity \emph{\cite{Sepulchre}}, which are all concepts that are closely related to notion of convergent systems \emph{\cite{Pav}} that we use here.
\end{remark}

\section{Problem Setup}\label{Setup}

\begin{figure*}[t]
  \centering
  \includegraphics[scale=.070]{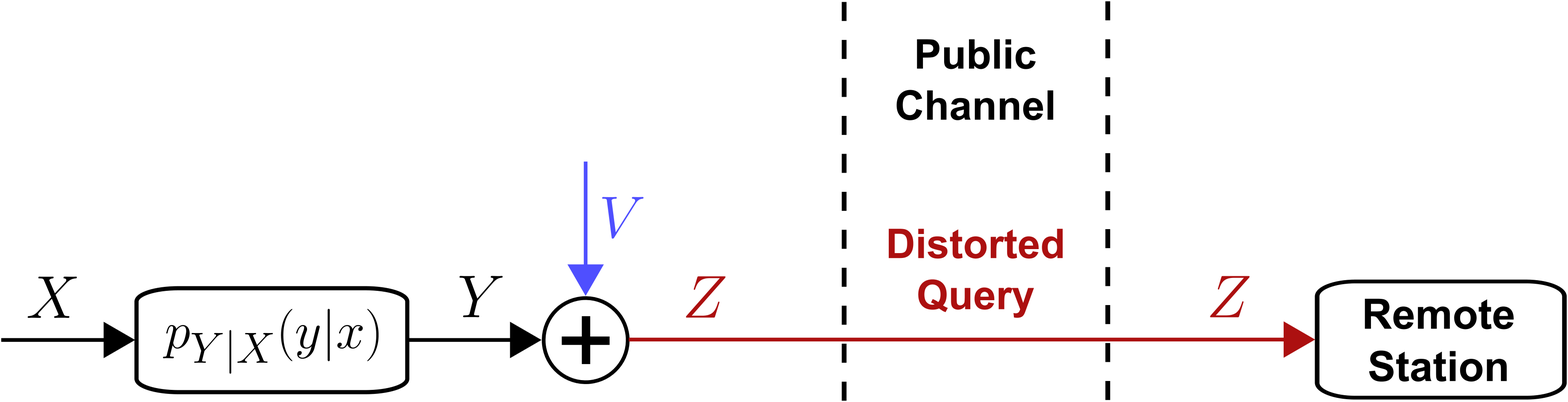}
  \caption{Configuration for Problem 1.}\label{Fig1b}
\end{figure*}

Let $X$ be a discrete random vector that must be kept private. The alphabet and probability mass function of $X$ are denoted as $\mathcal{X}=\{x_1,\ldots,x_{N}\}$, $x_i \in \Real^{n_x}$, $n_x \in \Nat$, $i\in \{1,\ldots,N\}$ and $p_X(x) = \text{Pr}[X=x]$, $x \in \mathcal{X}$, respectively. The $n_x$ entries of $X$ represent, for instance, private entries of $n_x$ users within a dataset that is stored by a trusted server. The server admits queries of the form $Y =q(X)$, $Y \in \Real^{n_y}$, for some (stochastic or deterministic) mapping $q:\Real^{n_x} \rightarrow \Real^{n_y}$ characterized by the transition probabilities $p_{Y|X}(y|x)$, $x \in \mathcal{X}$, $y \in \mathcal{Y}$, where $\mathcal{Y}=\{y_1,\ldots,y_{M}\}$, $y_i \in \Real^{n_y}$, $n_y \in \Nat$. The aim of our privacy scheme is to respond to queries of the form $q(X)$ with distorted queries $Z = q(X) + V$, for some discrete random vector $V$ (with $V \indep Y$), such that, when releasing $Z$, the individual entries of $X$ are ``hidden''. Realizations of the vector $Z$ are transmitted over a public (unsecured) communication channel to a remote station, see Figure \ref{Fig1b}. Then, if we do not add $V$ to $q(X)$ before transmission, information about $X$ is directly accessible through the public channel. As a preliminary problem that we need to solve for the subsequent results, we address the design of the probability distribution of $V$ to maximize privacy, i.e., the distribution of $V$ must be constructed so that the sum, $Z = q(X) + V$, carries as little information about $X$ as possible. In this manuscript, we use the \emph{mutual information} between $X$ and $Z = Y + V$, $I[X;Z]$, as \emph{privacy metric}. We aim at finding the probability mass function of $V$, $p_V(v)$, that minimizes $I[X;Z]$ over a class of probability mass functions. That is, we cast the design of $p_V(v)$ as an optimization problem with cost function $I[X;Z]$, optimization variables $p_V(v)$, and subject to $V \indep Y$ and the usual probability simplex constraints. Note that, contrary to related work \cite{Fawaz}-\nocite{Carlos_Iman1}\cite{Dullerud},\cite{Fawaz2},\cite{Poor}, we do not consider any sort of privacy-distortion trade-off in our formulation. We minimize $I[X;Y + V]$ regardless of the distortion between $Y$ and $Y + V$ induced by $V$. Distortion is not an issue because, we seek to generate exactly the same realization of $V$ at the remote station and then recover the query by subtracting this realization from the one of $Z=Y + V$. This is addressed in Problem 2 and Problem 3 below.

We let $V$ be a discrete random vector with alphabet $\mathcal{Y}$ and probability mass function $p_V(v) = \text{Pr}[V=v]$, $v \in \mathcal{Y}$, i.e., the alphabet of $V$ and the one of the query $Y=q(X)$ are equal. Having equal alphabets imposes a tractable convex structure on the cost $I[X;Z]$ and reduces the optimization variables to the probabilities of each element of the alphabet. The case with arbitrary alphabet leads to a combinatorial optimization problem where the objective changes its structure for different combinations. We do not address this case in this manuscript; it is left as a future work. In what follows, we formally present the optimization problem we seek to address.

\begin{problem}{\textbf{\emph{[Optimal Distribution of the Additive Distorting Signal]}}}
For given $p_X(x) = \text{\emph{Pr}}[X=x]$ and $p_{Y|X}(y|x) = \text{\emph{Pr}}[Y=y|X=x]$, $x \in \mathcal{X}$, $y \in \mathcal{Y}$, find the probability mass function  $p_V(v) = \text{\emph{Pr}}[V=v]$, $v \in \mathcal{Y}$ solution of the optimization problem:
\begin{equation} \label{eq:convex_optimization}
\left\{\begin{aligned}
	&p_V^*(v) := \argmin_{p_V(v)}\ I[X;V+Y],\\
    &\hspace{2mm}\text{\emph{s.t. }}  V \indep Y \text{ \emph{and} }p_V(v) \in \text{ \emph{Simplex}}.\\
\end{aligned}\right.
\end{equation}
\end{problem}

\begin{figure*}[t]
  \centering
  \includegraphics[scale=.070]{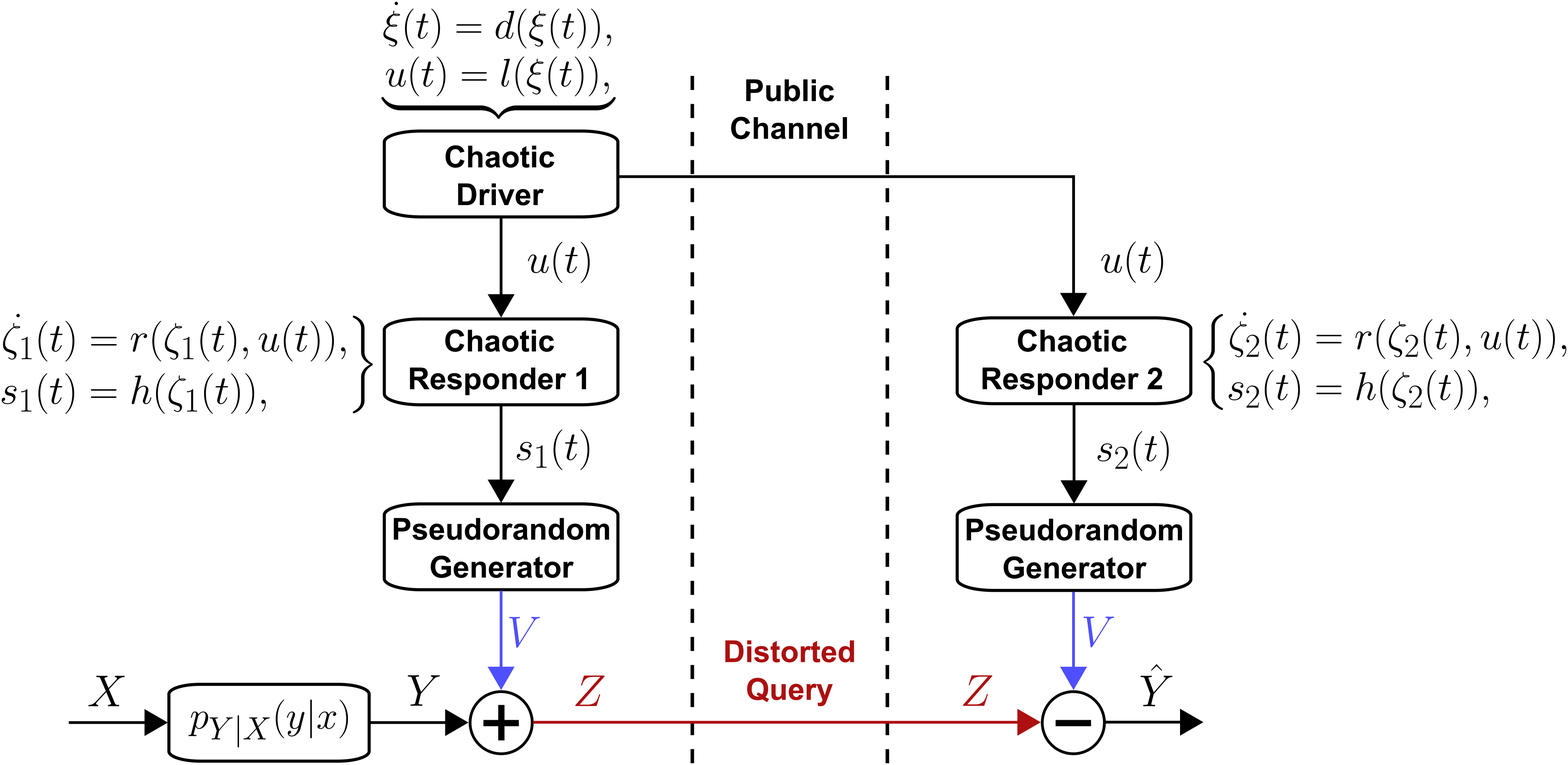}
  \caption{Complete System Configuration.}\label{Fig1}
\end{figure*}

Here, $p_V^*(v)$ denotes the optimal distribution solution to \eqref{eq:convex_optimization}. To hide $X$, once we have obtained $p_V^*(v)$, we aim at generating realizations $v \in \mathcal{Y}$ from this distribution, add them to the required query ($Y=q(X)$), and send realizations of the sum $Z=Y+V$ to the remote station through the public channel. At the other end of the channel, we seek to \emph{generate the exact same realizations from $p_V^*(v)$} so that we can recover the query by simply subtracting $V$ from $Z$, see Figure \ref{Fig1}. Note that, in Figure \ref{Fig1}, we have a recovered $\hat{Y}$ at the remote station rather that the actual $Y$. This is because we want to remark that, due to practical errors in our algorithm--e.g., due to communication delays and transients--realizations of $V$ that we generate at both ends of the channel might be slightly different in practice. To generate these realizations, we use trajectories, $\phi^\zeta_{u,1}(t,\zeta_1(0),u(t))$, $t \in \Real_{\geq 0}$, $\zeta_1(0) \in \Real^{n_\zeta}$, $u(t) \in \Real^{n_u}$, of a chaotic dynamical system of the form:
\begin{equation}
\left\{\begin{split}
\dot{\zeta_1}(t) &= r(\zeta_1(t),u(t)),  \label{2}\\
{s_1}(t) &= h(\zeta_1(t)),
\end{split}\right.
\end{equation}
with state $\zeta_1(t) \in \Real^{n_\zeta}$, output $s_1(t) \in \Real$, continuous in $t$ input $u(t) \in \mathcal{U} \subset \Real^{n_u}$ taking values in some compact set $\mathcal{U}$, continuous function $h:\Real^{n_\zeta} \rightarrow \Real$, and vector field $r:\Real^{n_\zeta} \times \mathcal{U} \rightarrow \Real^{n_\zeta}$ continuously differentiable in its first argument, uniformly in its second argument. Hereafter, system \eqref{2} is referred to as \emph{responder} 1. Responder 1 is placed at the side of the trusted server, see Figure \ref{Fig1}. The input signal $u(t)$ is generated by a chaotic autonomous exosystem:
\begin{equation}
\left\{\begin{split}
\dot{\xi}(t) &= d(\xi(t)),  \label{3}\\
{u}(t) &= l(\xi(t)),
\end{split}\right.
\end{equation}
with state $\xi(t) \in \Real^{n_\xi}$, output $u(t) \in \mathcal{U} \subset \Real^{n_u}$, and vector fields $d:\Real^{n_\xi} \rightarrow \Real^{n_\xi}$ and $l:\Real^{n_\xi} \rightarrow \Real^{n_u}$. The vector field $d(\xi)$ is locally Lipschitz in $\xi$ and $l(\xi)$ is continuous. We refer to \eqref{3} as the \emph{driver system}. We let $u(t)$ be connected to the remote station via the public channel, see Figure \ref{Fig1}. At the other end of the channel, driven by the same input signal $u(t)$, we have a third chaotic oscillator with the same dynamics as \eqref{2} but with potentially different initial conditions, i.e., the second oscillator is given by
\begin{equation}
\left\{\begin{split}
\dot{\zeta_2}(t) &= r(\zeta_2(t),u(t)),  \label{4}\\
{s_2}(t) &= h(\zeta_2(t)),
\end{split}\right.
\end{equation}
with state $\zeta_2(t) \in \Real^{n_\zeta}$ and output $s_2(t) \in \Real$. We denote trajectories of \eqref{4} as $\phi^\zeta_{u,2}(t,\zeta_2(0),u(t))$ with $t \in \Real_{\geq 0}$, $\zeta_2(0) \in \Real^{n_\zeta}$, and $u(t) \in \mathcal{U} \subset \Real^{n_\zeta}$. System \eqref{4} is referred to as \emph{responder} 2. Note that if $\zeta_1(t) = \zeta_2(t)$, $t \in \Real_{\geq 0}$, i.e., if systems \eqref{2} and \eqref{4} are synchronized, and we use the synchronous chaotic solution, say $\phi^\zeta_{u}(t,u(t))$, to generate realizations from $p_V^*(v)$, we could have the same realization of $V$ at both sides of the channel.

\begin{problem}{\textbf{\emph{[Boundedness, Chaos, and Synchronization]}}}\label{sync_problem}
State sufficient conditions on the vector fields $r(\cdot)$, $h(\cdot)$, $d(\cdot)$, and $l(\cdot)$ of the coupled system \eqref{2}-\eqref{4} such that: \emph{1)} trajectories of \eqref{2}-\eqref{4} exist and are bounded and chaotic; and \emph{2)} systems \eqref{2} and \eqref{4} exponentially synchronize, i.e., $\lim_{t \rightarrow \infty} | \zeta_1(t) - \zeta_2(t) |= 0$, exponentially fast.
\end{problem}

\begin{remark}\label{exponential_sync}
Problem \ref{sync_problem} seeks to enforce exponential synchronization by selecting the dynamics of the oscillators. Exponential synchronization implies that trajectories of the responders converge to each other exponentially for all initial conditions and are perfectly synchronized in the limit only. It follows that, in finite time, there is always a ``small'' difference between their trajectories. Nevertheless, because oscillators synchronize exponentially fast, and it is often possible in practice to select initial conditions from a known compact set (known to both the trusted server and the remote station), it is safe to assume that the interconnected systems have been operating for sufficiently large time such that oscillators are practically synchronized, i.e., the synchronization error is so small that trajectories can be assumed to be equal. This is a standard assumption that is made in most, if not all, of the existing work on chaotic encryption based on synchronization \emph{\cite{ChuaChao1}-\nocite{GRZYBOWSKI20092793}\nocite{LU2002365}\nocite{ALVAREZ2005775}\cite{ChuaChao2}}.
\end{remark}

Finally, once we have found functions solution to Problem 2, which guarantees exponential synchronization of the responders, and assuming that responders are synchronized (see Remark \ref{exponential_sync}), we aim at designing a procedure to generate pseudorandom realizations from $p_V^*(v)$ using the synchronous chaotic solution $\phi^\zeta_{u}(t,u(t))$. Note that $\zeta_1(t) = \zeta_2(t) \Rightarrow s_1(t) = h(\zeta_1(t)) = s_2(t) = h(\zeta_2(t))$, for all $t \geq 0$. Moreover, because $\zeta_1(t) = \zeta_2(t) = \phi^\zeta_{u}(t,u(t))$; then, $s_1(t) = s_2(t) = h(\phi^\zeta_{u}(t,u(t))) =: \phi^s_{u}(t,u(t)) \in \mathcal{S} \subset \Real$ for some compact set $\mathcal{S}$.\linebreak To reduce the complexity of the algorithm, we use the lower dimensional synchronous solution $\phi^s_{u}(t,u(t))$ to generate the realizations from $p_V^*(v)$.

\begin{problem}{\textbf{\emph{[Generation of Optimal Pseudorandom Numbers]}}}
Using the lower dimensional synchronous solution, $\phi^s_{u}(t,u(t))$, design an algorithm to generate pseudorandom realizations from the optimal distribution $p_V^*(v)$, $v \in \mathcal{Y}$.
\end{problem}

\section{Optimal Distribution of the Additive Distorting Signal}

In this section, we prove that Problem 1 can be posed as a convex program in the probabilities $p_V(v)$, $v \in \mathcal{Y}$. We derive an explicit expression for the cost function $I[X;Z]$, $Z=Y+V$, in terms of the given $p_X(x)$ and $p_{Y|X}(y|x)$ and the variables $p_V(v)$, restricted to satisfy the independence constraint $V \indep Y$.

\begin{lemma}\label{lem2}
$I[X;Z]$ with $Z=Y+V$, $V \indep Y$, is a convex function of $p_V(v)$, $v \in \mathcal{Y}$, for given $p_X(x)$ and $p_{Y|X}(y|x)$, $x \in \mathcal{X}$, $y \in \mathcal{Y}$; and can be written compactly in terms of $p_X(x)$, $p_{Y|X}(y|x)$, and $p_V(v)$, as follows:
\begin{subequations}
\begin{empheq}[left=\empheqlbrace]{align}
I[X;Z] &= \sum_{x \in \mathcal{X}}\sum_{z \in \mathcal{Z}}p_X(x)p_{Z|X}(z|x)\log \frac{p_{Z|X}(z|x)}{p_Z(z)},\label{cost1a}\\[1mm]
p_{Z|X}(z|x) &= \sum_{y \in \mathcal{Y}}p_{Y|X}(y|x)p_V(z-y),\label{cost1b}\\[1mm]
p_{Z}(z) &= \sum_{y \in \mathcal{Y}}p_{Y}(y)p_V(z-y).\label{cost1c}
\end{empheq}
\end{subequations}
\emph{\textbf{\emph{Proof}}: The expression on the right-hand side of \eqref{cost1a} follows by inspection of Definition \ref{mutual_info} and the fact that $p_{Z,X}(z,x) = p_X(x)p_{Z|X}(z|x)$. By \cite[Theorem 2.7.4]{Cover}, cost \eqref{cost1a} is convex in $p_{Z|X}(z|x)$ for given $p_X(x)$. However, our optimization variables are $p_V(v)$ and not $p_{Z|X}(z|x)$. Note that $X$, $Y$, and $Z$ form a Markov chain in that order \cite{Ross}; therefore, $p_{X,Y,Z}(x,y,z) = p_{X}(x)p_{Y|X}(y|x)p_{Z|Y}(z|y)$. Marginalizing $p_{X,Y,Z}(x,y,z)$ with respect to $Y \in \mathcal{Y}$ and then conditioning with respect to $X$ yields $p_{X,Z}(x,z) = \sum_{y \in \mathcal{Y}}p_{X}(x)p_{Y|X}(y|x)p_{Z|Y}(z|y)$ and $p_{Z|X}(z|x) = \sum_{y \in \mathcal{Y}}p_{Y|X}(y|x)p_{Z|Y}(z|y)$, respectively. Note that $p_{Z|X}(z|x)$ is just a linear transformation of $p_{Z|Y}(z|y)$. Hence, convexity with respect to $p_{Z|X}(z|x)$ implies convexity with respect to $p_{Z|Y}(z|y)$ because convexity is preserved under affine transformations \cite{Boyd2004}. Next, consider $p_{Z|Y}(z|y) = p_{Z,Y}(z,y)/p_{Y}(y)$. By definition, $p_{Z,Y}(z,y) = \text{Pr}[Z=z,Y=y]$, $z \in \mathcal{Z}$, $y \in \mathcal{Y}$. Note that
\begin{align*}
\text{Pr}[Z=z,Y=y] &= \text{Pr}[Y+V=z,Y=y] = \text{Pr}[V=z-y,Y=y]\\
                   &\overset{\text{(a})}{=} \text{Pr}[V=z-y]\text{Pr}[Y=y] = p_V(z-y)p_Y(y),
\end{align*}
where (a) follows from independence between $V$ and $Y$. Thus,
\begin{align*}
p_{Z|Y}(z|y) &= \frac{p_{Z,Y}(z,y)}{p_{Y}(y)}\\
             &= \frac{p_V(z-y)p_Y(y)}{p_{Y}(y)} = p_V(z-y).
\end{align*}
We have concluded convexity of $I[X;Z]$ with respect to $p_{Z|Y}(z|y)$ above. Hence, because $p_{Z|Y}(z|y) = p_V(z-y)$ and $p_V(z-y)$ is a linear transformation of $p_V(v)$ ($p_V(z-y) = p_V(v)$ for $z-y=v$ and zero otherwise), the cost $I[X;Z]$ is convex in $p_V(v)$. Moreover, since $p_{Z|X}(z|x) = \sum_{y \in \mathcal{Y}}p_{Y|X}(y|x)p_{Z|Y}(z|y)$ and $p_{Z|Y}(z|y) = p_V(z-y)$, equality \eqref{cost1b} holds true. It remains to prove that $p_Z(z)$ can be written as \eqref{cost1c}. Because $Z=Y+V$, $p_Z(z) = \text{Pr}[Z=z]$, for a given $z \in \mathcal{Z}$, can be written as the sum of the probabilities of all $Y=y$ and $V=v$ that result in $z$, i.e.,
\begin{align*}
p_Z(z) &= \text{Pr}[Z=z] = \text{Pr}[Y+V=z]\\
                   &= \sum_{y \in \mathcal{Y}}\text{Pr}[V=z-y,Y=y]\\
                   &\overset{(b)}{=} \sum_{y \in \mathcal{Y}}\text{Pr}[Y=y]\text{Pr}[V=z-y] = \sum_{y \in \mathcal{Y}}p_Y(y)p_{V}(z-y),
\end{align*}
where (b) follows from independence between $V$ and $Y$. \hfill $\blacksquare$
}\end{lemma}
By Lemma \ref{lem2}, the cost $I[X;Z]$, for $V \indep Y$, is convex in $p(v)$ and parametrized by $p_X(x)$ and $p_{Y|X}(y|x)$. In what follows, we cast the nonlinear program for solving Problem 1.\\

\begin{theorem}\label{th2}
Given $p_X(x)$ and $p_{Y|X}(y|x)$, $x \in \mathcal{X}$, $y \in \mathcal{Y}$, the mapping $p_V(v)$, $v \in \mathcal{Y}$, that minimizes $I[X;Z]$, $Z = V + Y$, subject to $V \indep Y$ can be found by solving the following convex program:
\begin{equation} \label{eq:convex_optimization4}
\left\{\begin{aligned}
	&p_V^*(v) = \argmin_{p_V(v)} \sum_{x \in \mathcal{X}}\sum_{y \in \mathcal{Y}}\sum_{z \in \mathcal{Z}}p_X(x)p_{Y|X}(y|x)p_V(z-y)\log \frac{\sum_{y \in \mathcal{Y}}p_{Y|X}(y|x)p_V(z-y)}{\sum_{y \in \mathcal{Y}}p_{Y}(y)p_V(z-y)},\\[2mm]
    &\hspace{15mm}\text{ \emph{s.t. }} p_V(v) \in \text{ \emph{Simplex}}.
\end{aligned}\right.
\end{equation}
\emph{\textbf{\emph{Proof}}: Theorem 1 follows from Lemma \ref{lem2}.}
\end{theorem}

\section{Boundedness, Chaos, and Synchronization}

\subsection{Existence, Uniqueness, and Boundedness of Solutions}

We start addressing existence, uniqueness, and boundedness of the solutions of the coupled systems \eqref{2}-\eqref{4}. To be able to use synchronous solutions to generate realizations from $p_V^*(v)$, we first need these solutions to exist and be bounded. In the system description given above, we have assumed that $r(\zeta,u(t))$ is continuously differentiable in $\zeta$ uniformly in $u(t)$, $u(t)$ is continuous in $t$, and $d(\xi)$ is locally Lipschitz. These alone imply uniqueness and existence of solutions of \eqref{2}-\eqref{4} over some finite time interval $t \in [0,\tau]$, $\tau \in \Real_{>0}$, \cite[Theorem 2.2]{Kha02}.\linebreak To conclude the latter for arbitrarily large $\tau$, besides the locally Lipschitz assumption on the functions, we need boundedness of the solutions of \eqref{2}-\eqref{4} \cite[Theorem 2.4]{Kha02}. Note that the coupled systems \eqref{2}-\eqref{4} have a cascade structure. The driver dynamics is independent of the responders states, and its output, $u(t)$, is the input of the responders. Then, an approach to conclude boundedness of the overall system is to conclude boundedness of the driver first, and then boundedness of the responders when driven by bounded inputs. In what follows, we formally introduce the notion of boundedness that we use here.

\begin{definition}\emph{\cite{Kha02}}\label{boundedness}
The solutions of \eqref{3} are bounded for a bounded set of initial conditions if there exists a positive constant $c$, independent of the initial time instant, and for every $a \in (0,c)$, there is $b = b(a) > 0$, independent of the initial time instant, such that $|\xi(0)|\leq a \Rightarrow |\xi(t)|\leq b$, $\forall \hspace{1mm} t \geq 0$. If the latter holds for arbitrarily large $a$; then, the solutions of \eqref{3} are globally bounded.
\end{definition}

\begin{remark}
Because $l(\xi)$ is continuous, by the extreme value theorem, boundedness of $\xi(t)$ implies boundedness of $u(t) = l(\xi(t))$.
\end{remark}

\begin{remark}\label{BoundedCOnd}
We do not give conditions for boundedness of the solutions of \eqref{3}. It is assumed that the vector field $d(\xi)$ is such that the solutions of the driver are globally bounded. We refer the reader to, for instance, \linebreak \emph{\cite[Theorem 4.18]{Kha02}}, where sufficient conditions for boundedness are given in terms of Lyapunov-like results.
\end{remark}

Next, for bounded solutions of the driver, we need the solutions of the responders to be bounded when driven by $u(t)$. To address this, we use the notion if Input-to-State-Stability (ISS) \cite{SONTAG}.

\begin{definition}\emph{\cite{SONTAG}}\label{ISS2}
System \eqref{2} (and thus system \eqref{3} as well) is said to be Input-to-State-Stable if there exist a class $\mathcal{KL}$ function $\beta(\cdot)$ and a class $\mathcal{K}$ function $\gamma(\cdot)$ such that for any initial condition $\zeta_1(0)$ and any bounded input $u(t)$, the solution $\zeta_1(t)$ exists for all $t \in \Real_{\geq 0}$ and satisfies: $|\zeta_1(t)| \leq \beta(\zeta_1(0),t) + \gamma\left( ||u||_\infty \right)$.
\end{definition}

\begin{remark}
\emph{ISS} of the responders with respect to $u(t)$ guarantees that, for any bounded $u(t)$, the states $\zeta_1(t)$ and $\zeta_2(t)$ are bounded. Moreover, as $t$ increases, $|\zeta_1(t)|$ and $|\zeta_2(t)|$ are ultimately bounded \emph{\cite{Kha02}} by $\gamma\left( ||u||_\infty \right)$, see \emph{\cite{SONTAG}} for further details.
\end{remark}

\begin{remark}\label{ISS}
Sufficient conditions for the responders to be ISS with input $u(t)$ are not provided here. We assume that the vector field $r(\zeta,u(t))$ is such that systems \eqref{2} and \eqref{3} are ISS with respect to $u(t)$. We refer the reader to, for instance, \emph{\cite[Theorem 4.19]{Kha02}}, where sufficient conditions for ISS are given in terms of ISS-Lyapunov functions.
\end{remark}

\begin{remark}\label{iISS}
The weaker property of integral Input-to-State-Stable \emph{(iISS)} \emph{\cite{Arcak02aunifying}} could be used to conclude boundedness of the responder's trajectories when driven by ``sufficiently small'' inputs. We refer the reader to \emph{\cite{AngeliiISS}}, where sufficient conditions for \emph{iISS } and related boundedness results are given.
\end{remark}

\subsection{Synchronization}

Next, we give sufficient conditions on $r(\zeta,u(t))$ such that $\lim_{t \rightarrow \infty} | \zeta_1(t) - \zeta_2(t) | = 0$, i.e., the responders exponentially synchronize. We assume that solutions of the coupled systems \eqref{2}-\eqref{4} exist and are bounded, i.e., vector fields $r(\cdot)$, $d(\cdot)$, and $l(\cdot)$ satisfy the conditions stated in the previous subsection. Then, for bounded $u(t)$, a sufficient condition for the responders to exponentially synchronize is that systems \eqref{2} and \eqref{4} are \emph{convergent systems} in the sense of Definition \ref{converg}. The latter implies that, because both responders are driven by the input $u(t)$ and their dynamics are described by the same set of differential equations, trajectories of \eqref{2} and \eqref{4} converge to the same the limit solution, $\phi^\zeta_{u}(t,u(t))$, and this solution is solely determined by $u(t)$ and not by the initial conditions. In the following corollary of Proposition \ref{converg2}, we give a sufficient condition for the responders to be exponentially convergent (and thus to exponentially synchronize).

\begin{corollary}\label{cor1}
Consider the responders \eqref{2} and \eqref{4}. If there exists a positive definite matrix $P \in \Real^{n_\zeta \times n_\zeta}$ such that, for all $u \in \Real^{n_{u}}$ and $\zeta \in \Real^{n_{\zeta}}$, all the eigenvalues of the symmetric matrix:
\begin{equation}
\frac{1}{2} \left( P\left( \frac{\partial r}{\partial \zeta }%
(\zeta,u)\right) +\left( \frac{\partial r}{\partial \zeta }(\zeta
,u)\right) ^{T} P \right),   \label{converResponders}
\end{equation}
are negative and separated from zero; then, responders \eqref{2} and \eqref{4} are globally exponentially convergent, and thus $\lim_{t \rightarrow \infty} |\zeta_1(t) - \zeta_2(t)| = 0$, exponentially fast.
\end{corollary}

\begin{remark}\label{quantization}
If the driver's output $u(t)$ is to be sent over a network and quantization (or some sort of coding) is required, we would need to drive responders by the same quantized $u(t)$, say $u_Q(t)$, to achieve exponential synchronization. That is, if we quantize $u(t)$ to obtain $u_Q(t)$, and we drive both responders by $u_Q(t)$ (with, e.g., a Zero-Order-Hold \emph{(ZOH)}), they would also exponentially synchronize. They would synchronize to a different trajectory than when driven by $u(t)$, but they would synchronize exponentially fast.
\end{remark}

Besides the notion of convergent systems, there are other methods available in the literature that can be used to verify that trajectories of responders asymptotically synchronize to a limit solution that is independent of the initial conditions. See Remark \ref{incrementalISS} for details.\\

\subsection{Chaotic Dynamics}\label{chaos}

There are mainly two branches of methods to identify chaotic dynamics; namely, standard largest Lyapunov exponent methods \cite{wiggins}, and the more recent (0-1) test \cite{Melbourne}. Both methods use trajectories (numerical or experimental) of the systems under study to decide whether they are chaotic or not. In general, there are no sufficient conditions directly on the differential equations (the vector fields $r(\cdot)$ and $d(\cdot)$) such that chaotic trajectories are guaranteed to occur. There are, however, many well known systems in the literature known to exhibit chaotic trajectories. For instance, the Lorenz system \cite{strogatz:2000}, Duffing \cite{kovacic2011duffing} and van der Pol \cite{VanderPol} oscillators, the R\"{o}ssler \cite{Pecora} and Chua \cite{Chua} systems, and neural oscillators \cite{Erik3} (e.g., the Hodgkin-Huxley, Morris-Lecar, Hindmarsh-Rose, and FitzHugh-Nagumo oscillators). We can use any of these chaotic systems (if they satisfy all the required extra conditions, see Section \ref{synth}) as driver and then select a pair of responders with convergent dynamics. Indeed, we need to verify that the responders that we choose produce chaotic trajectories when driven by the chaotic driver. Moreover, to generate the pseudorandom realizations from $p_V^*(v)$ (this is addressed in the next section), we need the chaotic trajectories of the responders, regarded as a random process, to be \emph{stationary}, i.e., after transients have settled down, trajectories must follow a stationary probability distribution \cite{Ross} which is independent of the initial conditions. The latter is a strong condition that is not satisfied for all chaotic systems. The existence of stationary distributions for chaotic trajectories has been proven for hyperbolic and quasi-hyperbolic (also called singular-hyperbolic) chaotic systems \cite{Anishchenko}. The definition of (quasi) hyperbolic dynamical systems \cite{Anishchenko,kuznetsov2012hyperbolic} is technical and not needed for the subsequent results. It requires concepts from differential topology that we prefer to omit here for readability of the manuscript. It suffices to know that the chaotic system that we use for the driver must lead to stationary distributions of the responders. This can be tested numerically by Monte Carlo simulations \cite{Robert:2005:MCS:1051451}. Moreover, there are many well-known chaotic systems with (quasi) hyperbolic dynamics in the literature, e.g., the Lorenz and Chua systems \cite{Kapitaniak}, neural oscillators \cite{Belykh2005HyperbolicPA}, the many predator-pray like systems given in \cite{Kuznetsov2007AutonomousCO,TURUKINA20111407}, and some mechanical nonlinear oscillators \cite{Kuznetsov2017}. In the next subsection, we provide a synthesis procedure to choose the functions of the coupled systems \eqref{2}-\eqref{4} such that all the required conditions mentioned above are satisfied.\\

\subsection{General Guidelines}\label{synth}

\noindent\rule{\hsize}{1pt}\vspace{.2mm}
\textbf{Synthesis Procedure:}\\[.5mm]
\textbf{1)} Select a driver dynamics \eqref{3} (i.e., the vector field $d(\xi)$) known to be chaotic and (quasi) hyperbolic (e.g., systems in \cite{Kapitaniak}-\nocite{Belykh2005HyperbolicPA}\nocite{Kuznetsov2007AutonomousCO,TURUKINA20111407}\cite{Kuznetsov2017}).\\[.5mm]
\textbf{2)} Verify that the corresponding $d(\xi)$ is locally Lipschitz and the trajectories of the driver are globally bounded,
in the sense of Definition \ref{boundedness}, using, e.g., \cite[Theorem 4.18]{Kha02}.\\[.5mm]
\textbf{3)} In \eqref{3}, let $\xi= (\xi^1,\ldots,\xi^{n_\xi})^\top \in \Real^{n_\xi}$, $\xi^i \in \Real$, and $u(t) = l(\xi(t)) = \xi^j(t)$, $i,j \in \{ 1,\ldots,n_\xi \}$, i.e., fix the output
of the driver to be any state of \eqref{3}. In doing this, we ensure that $u(t)$ is continuous, bounded, chaotic, and (quasi) hyperbolic.\\[.5mm]
\textbf{4)} For the responders \eqref{2} and \eqref{4}, select any continuously differentiable vector field $r(\zeta,u)$ (with respect to $\zeta$) leading to
ISS dynamics, see Remark \ref{ISS}, and satisfying the conditions for convergence in Corollary \ref{cor1}, e.g., $r(\zeta,u) = A\zeta + \psi(u)$, for any matrix $A \in \Real^{n_\zeta \times n_\zeta}$ with spectral radius $\rho[A]<1$ and differentiable vector field $\psi:\Real^{n_u} \rightarrow \Real^{n_\zeta}$. Then, we ensure that the
 responders have bounded trajectories and exponentially synchronize.\\[.5mm]
\textbf{5)} Verify that the trajectories of the responders, when driven by the chaotic driver, are chaotic (using Lyapunov exponents or the (0-1) test) and, after transients have settled down, lead to a stationary probability distribution independent of the initial conditions. See Section \ref{chaos} for details.\\[.5mm]
\textbf{6)} In \eqref{2} (and respectively in \eqref{4}), let $\zeta_1= (\zeta_1^1,\ldots,\zeta_1^{n_\zeta})^\top \in \Real^{n_\zeta}$, $\zeta_1^i \in \Real$, and $s_1(t) = l(\zeta_1(t)) = \zeta_1^j(t)$, $i,j \in \{ 1,\ldots,n_\xi \}$, i.e., fix the output of the responders to be any state of \eqref{2} and \eqref{4}, respectively. Indeed, we need the same $j$ for both responders, i.e., $s_1(t) = \zeta_1^j(t)$ and $s_2(t) = \zeta_2^j(t)$. In doing this, we ensure that $s_1(t)$ and $s_2(t)$ are continuous, bounded, chaotic, and lead to stationary probability distributions.\\ \vspace{.2mm}\noindent\rule{\hsize}{1pt}

\section{Generation of Optimal Pseudorandom Numbers}\label{numbers}

In this section, we assume that the driver and the responders dynamics have been designed following the general guidelines in Section \ref{synth}. Then, for sufficiently large $t$, the chaotic trajectories of the responders are practically synchronized, i.e., for any finite $t^* \in \Real_{>0}$, there is $\epsilon_{t^*} \in \Real_{>0}$, such that $|s_1(t) - \phi^s_{u}(t,u(t))| \leq \epsilon_{t^*}$ and $|s_2(t) - \phi^s_{u}(t,u(t))| \leq \epsilon_{t^*}$, for all $t \geq t^*$, where $\phi^s_{u}(t,u(t)) \in \mathcal{S} \subset \Real$ denotes the asymptotic synchronous solution for some compact set $\mathcal{S}$; and samples from $\phi^s_{u}(t,u(t))$ follow a stationary probability distribution. Here, we assume that the responders have been operating for sufficiently large time such that the synchronization error, $|s_1(t)-s_2(t)|$, is so small that trajectories of the responders can be assumed to be equal to $\phi^s_{u}(t,u(t))$ (see Remark \ref{exponential_sync}), i.e., $t^*$ is sufficiently large so that $\epsilon_{t^*}$ is practically zero. In Section \ref{distortion}, we quantify the worst-case distortion induced by assuming $s_1(t)=s_2(t)=\phi^s_{u}(t,u(t))$ in finite time. In particular, we give an upper bound on the mean squared error $E[|Y-\hat{Y}|^2]$, where $\hat{Y}$ denotes the estimate of realizations of $Y$ using $s_1(t)$, $s_2(t)$, and the algorithm provided below. In the remainder of this section, we assume $s_1(t)=s_2(t)=\phi^s_{u}(t,u(t))$. Note that the sample space of $\phi^s_{u}(t,u(t))$, regarded as a random process, is some compact set $\mathcal{S} \subset \Real$, i.e., the sample space is a subset of the real line and thus samples from $\phi^s_{u}(t,u(t))$ follow some stationary probability density function (pdf), say $f_S(s)$, for some virtual continuous random variable $S$. That is, for $s(t) := \phi^s_{u}(t,u(t))$, define the sampled sequence $s_k := s(t_k)$ for sampling time-instants $t_k \in \Real_{>0}$, $t_k := \Delta k$, $k \in \Nat$, and sampling period $\Delta \in \Real_{>0}$; then, $s_k \sim f(s)$ for all $k$. Because we know the dynamics \eqref{2}-\eqref{4}, we can obtain $f_S(s)$ by Monte Carlo simulations \cite{Robert:2005:MCS:1051451}. If we know $f_S(s)$, we can always find a set of cells $C := \{c^1,\ldots,c^M \}$, $M \in \Nat$, $j \in \{1,\ldots,M\}$, such that $\bigcup_j c^j = \Real$, $\bigcap_j c^j = \emptyset$, and $\text{Pr}[s_k \in c] = \text{Pr}[V=v] = p_V^*(v)$ for $v \in \mathcal{Y}$ and $c \in C$. In other words, using the pdf $f_S(s)$, we can select the cells $C$ so that the probability that $s_k$ lies in the cells equals the optimal probability distribution $p_V^*(v)$. It follows that we can generate pseudorandom realizations from $p_V^*(v)$ by properly selecting $C$. Note that, because realizations are being generated by a deterministic process, there would be high correlation between consecutive realizations for small sampling period $\Delta$. However, because the $s_k$ is a stationary process (see Section \ref{chaos}), the larger the $\Delta$, the smaller the correlation between $s_{k}$ and $s_{k+1}$ for all $k \in \Nat$. Indeed, large $\Delta$ would introduce large time-delays for generating realizations. There is a trade-off between correlation and time-delay that should be taken into account in practice. One way to deal with this trade-off is to compute the normalized autocorrelation function \cite{Anishchenko,ChuaChao2} of $s_k$. Then, we select the smallest time-delay $\tau \in \Nat$ that leads to a desired correlation between $s_{k}$ and $s_{k+\tau}$, $k \in \Nat$, and use the delayed sequence $s^\tau(\cdot) := \{ s_{k},s_{k+\tau},s_{k+2\tau},\ldots \}$ to generate realizations from $p_V^*(v)$. In the following algorithm, we summarize the ideas introduced above.\\

\noindent\rule{\hsize}{1pt}\vspace{.2mm}
\textbf{Algorithm 1: Pseudorandom Number Generation:}\\[1mm]
\textbf{1)} Consider the probability mass function $p_V^*(v) = \text{Pr}[V=v]$, $v \in \mathcal{Y} = \{y_1,\ldots,y_M\}$, solution to Problem 1; and the synchronous solution $s(t) = \phi^s_{u}(t,u(t))$ of the responders.\\[1mm]
\textbf{2)} Fix the sampling period $\Delta \in \Real_{>0}$ and obtain, by Monte Carlo simulations \cite{Robert:2005:MCS:1051451}, the probability density function $f_S(s_k)$ of the sampled sequence $s_k = s(t_k)$, $t_k = \Delta k$, $k \in \Nat$.\\[1mm]
\textbf{3)} Select a finite set of cells $C = \{c^1,\ldots,c^M \}$, $M \in \Nat$, $j \in \{1,\ldots,M\}$, such that $\bigcup_j c^j = \Real$, $\bigcap_j c^j = \emptyset$, and $\text{Pr}[s_k \in c^j] = \text{Pr}[V=y_j]$ for all $y_j \in \mathcal{Y}$.\\[1mm]
\textbf{4)} Generate realization from $p_V^*(v)$ using the piecewise function:
\begin{equation} \label{piecewise}
v_k = \psi(s_k) :=
\small\left\{
\begin{array}{l}
y_1 $  \hspace{1.75mm}if $ s_k \in c^1, \\ \hspace{11mm} \vdots \\
y_M $ if $ s_k \in c^M.
\end{array}
\right.
\end{equation}
\\ \vspace{.2mm}\noindent\rule{\hsize}{1pt}\vspace{1mm}

\begin{figure}[t]
  \centering
  \includegraphics[scale=.1]{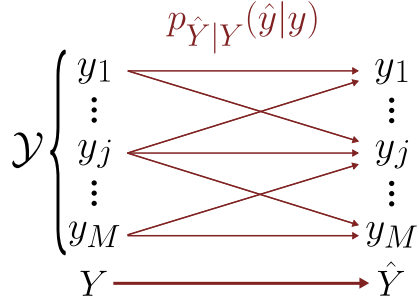}
  \caption{Transition probabilities $p_{\hat{Y}|Y}(\hat{y}|y)$.}\label{pmf}
\end{figure}

\subsection{Distortion Induced by Synchronization Errors}\label{distortion}

Algorithm 1 in Section \ref{numbers} is constructed under the assumption that responders are perfectly synchronized. However, because we only have exponential synchronization, in finite time, there is always a ``small'' difference between $s_1(t)$ and $s_2(t)$ due to potentially different initial conditions. It follows that there is also a difference between realizations generated using $s_1(t_k)$, denoted as $v_k^1 \in \mathcal{Y}$, and realizations $v_k^2 \in \mathcal{Y}$ generated through $s_2(t_k)$, where $\mathcal{Y} = \{ y_1,\ldots,y_M \}$. Exponential synchronization implies that for any finite $t^* \in \Real_{>0}$, there is $\delta(t^*,|s_1(0) - s_2(0)|) \in \Real_{>0}$ (denoted as $\delta_{t^*}$ for simplicity), parametrized by $t^*$ and the initial synchronization error $|s_1(0) - s_2(0)|$, such that $|s_1(t_k) - s_2(t_k)| \leq \delta_{t^*}$ for all $t_k \geq t^*_k$, and $\lim_{k \rightarrow \infty} |s_1(t_k) - s_2(t_k)| = 0$. Consider the cell $c^j$, $c^j \in C$, with end points $c^j_1$ and $c^j_2$, $c^j_1<c^j_2$, the length of $c^j$ is defined as $l(c^j):= c^j_2 - c^j_1$. If $c^j_1 = \pm \infty$ (or $c^j_2 = \pm \infty$), $l(c^j) = \infty$. Without loss of generality, let $l(c^2) \leq l(c^3) \leq \ldots \leq l(c^{M-1})$, $l(c^1) = \infty$, and $l(c^M) = \infty$. Note that, if $\delta_{t^*} \leq l(c^2)$, $v_k^1$ and $v_k^2$ are at most one level apart from each other, e.g., if $v_k^1 = y_1$, then either $v_k^2 = y_1$ or $v_k^2 = y_2$; and if $v_k^1 = y_3$, then $v_k^2 = y_2$, $v_k^2 = y_3$, or $v_k^2 = y_4$. It follows that $p_{\hat{Y}|Y}(\hat{y}|y)$, $y,\hat{y} \in \mathcal{Y}$, is of the form depicted in Figure \ref{pmf}, where $\hat{Y}$ denotes the estimate of realizations of $Y$ using $s_1(t_k)$, $s_2(t_k)$, and Algorithm 1. Similarly, if $l(c^2) < \delta_{t^*} \leq l(c^3)$, $v_k^1$ and $v_k^2$ are at most two levels apart from each other and thus lead to a different structure of the transition probabilities. Here, we only consider the case where $\delta_{t^*} \leq l(c^2)$. Distortion induced by larger synchronization errors can be estimated following the same methods. Note that, because responders synchronize exponentially, as $\delta_{t^*} \rightarrow 0$ ($t^* \rightarrow \infty$), $p_{\hat{Y}|Y}(\hat{y}|y) \rightarrow 1$ for $\hat{y}=y$, and $p_{\hat{Y}|Y}(\hat{y}|y) \rightarrow 0$, for $\hat{y} \neq y$, for all $y,\hat{y} \in \mathcal{Y}$. That is, distortion due to synchronization errors disappears exponentially fast. The actual value of the transition probabilities depend on the responders and driver dynamics, the initial conditions, and the cells $C$. However, we do not need these probabilities, only the structure of $p_{\hat{Y}|Y}(\hat{y}|y)$ depicted in Figure \ref{pmf} is used to derive an upper bound on the expected distortion. Let $\mathcal{V}_{\delta} \subseteq \mathcal{Y} \times \mathcal{Y}$ denote the set of pairs $(y_j,y_i)$ for which there is a nonzero transition probability $p_{\hat{Y}|Y}(y_j|y_i)$ between $Y = y_j$ and $\hat{Y} = y_i$, $y_j,y_i \in \mathcal{Y}$, as depicted in Figure \ref{pmf}. The set $\mathcal{V}_{\delta}$ is parametrized by the upper bound on the synchronization error $|s_1(t_k) - s_2(t_k)| \leq \delta_{t^*} \leq l(c^2)$. Define the distortion function $d(Y,\hat{Y}):= |Y - \hat{Y}|^2$. The function $d(Y,\hat{Y})$ is a deterministic function of two jointly distributed random vectors, $Y$ and $\hat{Y}$, with joint distribution $p_{Y,\hat{Y}}(y,\hat{y}) = p_{Y}(y)p_{\hat{Y}|Y}(\hat{y}|y)$. Hence, see \cite{Ross} for details, we can write the expected distortion as follows
\begin{align}
E[d(Y,\hat{Y})] &= \sum_{y,\hat{y} \in \mathcal{Y}}p_{Y,\hat{Y}}(y,\hat{y})d(y,\hat{y}) = \sum_{y,\hat{y} \in \mathcal{Y}}p_{Y}(y)p_{\hat{Y}|Y}(\hat{y}|y)|y - \hat{y}|^2 \notag\\
                &= \sum_{(y,\hat{y}) \in \mathcal{V}_{\delta}}p_{Y}(y)p_{\hat{Y}|Y}(\hat{y}|y)|y - \hat{y}|^2 \leq \sum_{(y,\hat{y}) \in \mathcal{V}_{\delta}}p_{Y}(y)|y - \hat{y}|^2 =: \bar{d}_{\delta},\label{distor2}
\end{align}
where the left-hand side of \eqref{distor2} follows from the definition of $\mathcal{V}_{\delta}$ above, and the last inequality from the fact that $p_{\hat{Y}|Y}(\hat{y}|y) \leq 1$ for all $y,\hat{y} \in \mathcal{Y}$. The constant $\bar{d}_{\delta} \in \Real_{>0}$ provides an upper bound on the worst-case distortion induced by a $\delta_{t^*}$ synchronization error. Moreover, as $\delta_{t^*} \rightarrow 0$, $\mathcal{V}_{\delta} \rightarrow \{(y_1,y_1),(y_2,y_2),\ldots,(y_M,y_M)\}$; therefore, $\lim_{\delta_{t^*} \rightarrow 0}\bar{d}_{\delta} = 0$. That is, distortion due to synchronization errors is bounded by $\bar{d}_{\delta}$ and vanishes exponentially fast.

\section{Simulation Results}

\begin{figure}[t]
  \centering
  \includegraphics[scale=.3]{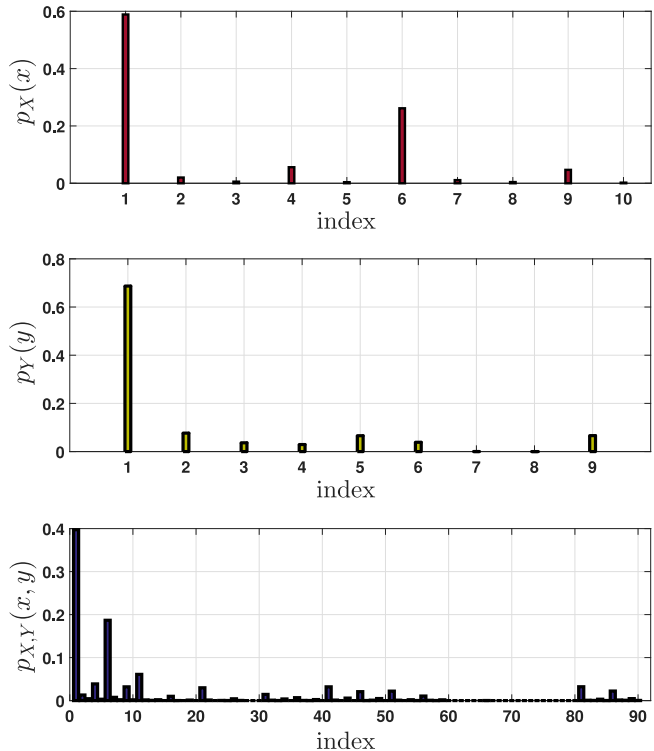}
  \caption{Probability mass functions of $X$, $Y$, and $(X,Y)$.}\label{Fig3}
\end{figure}

\begin{figure}[t]
  \centering
  \includegraphics[scale=.3]{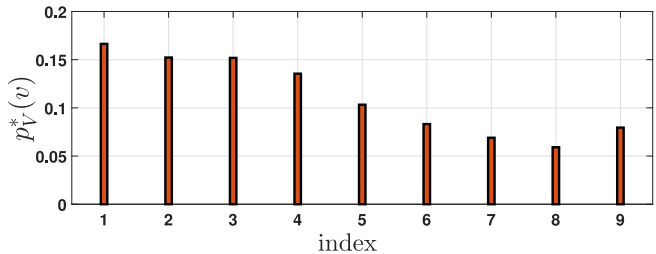}
  \caption{Optimal distribution $p_V^*(v)$ solution to \eqref{eq:convex_optimization4} in Theorem \ref{th2}.}\label{Fig_opt}
\end{figure}

\begin{figure}[t]
  \centering
  \includegraphics[scale=.275]{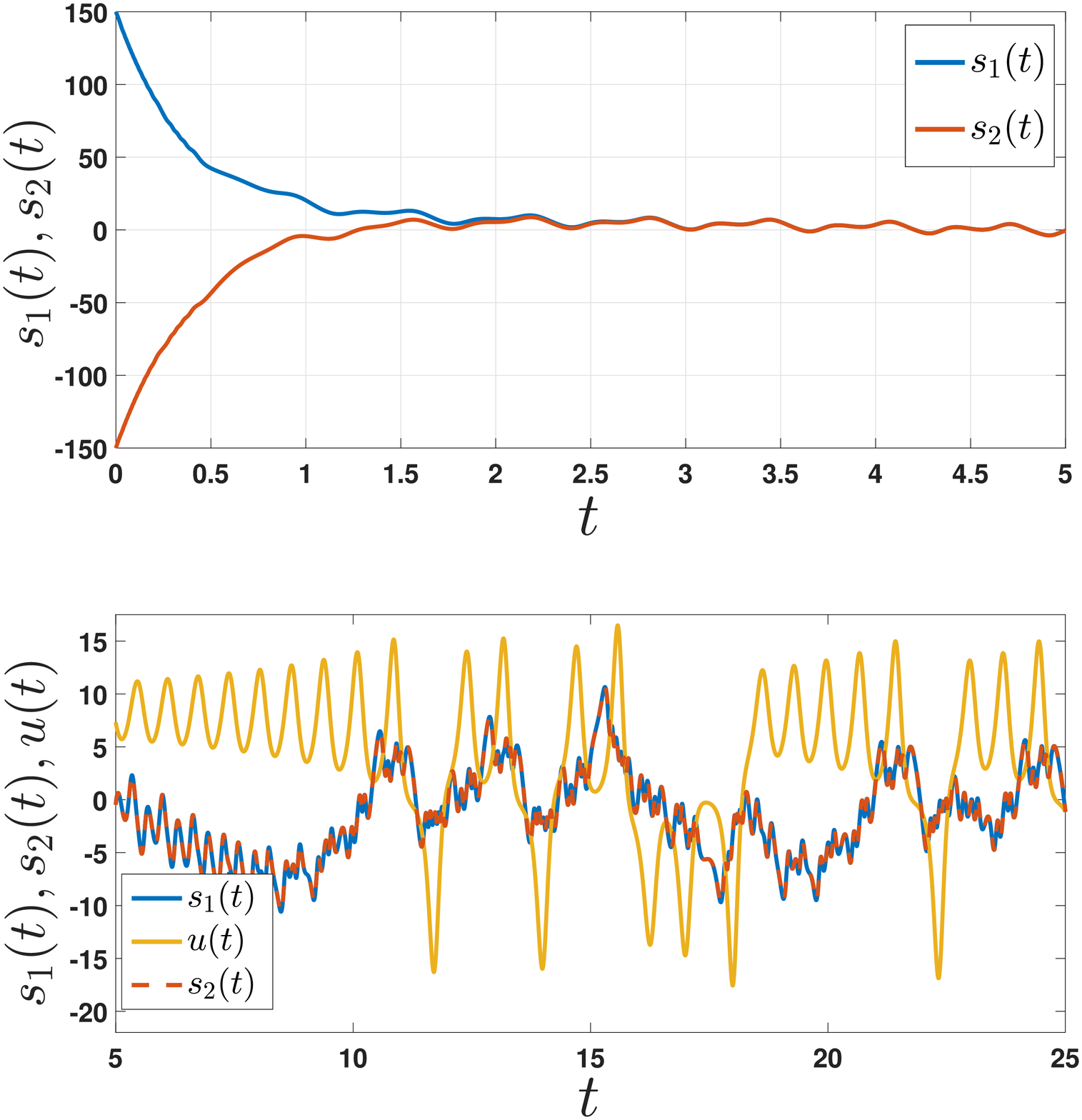}
  \caption{Traces of the chaotic driver and responders trajectories. Top: trajectories of the responders converging to each other. Bottom: traces of chaotic solutions of the driver and responders.}\label{Fig4}
\end{figure}

\begin{figure}[t]
  \centering
  \includegraphics[scale=.275]{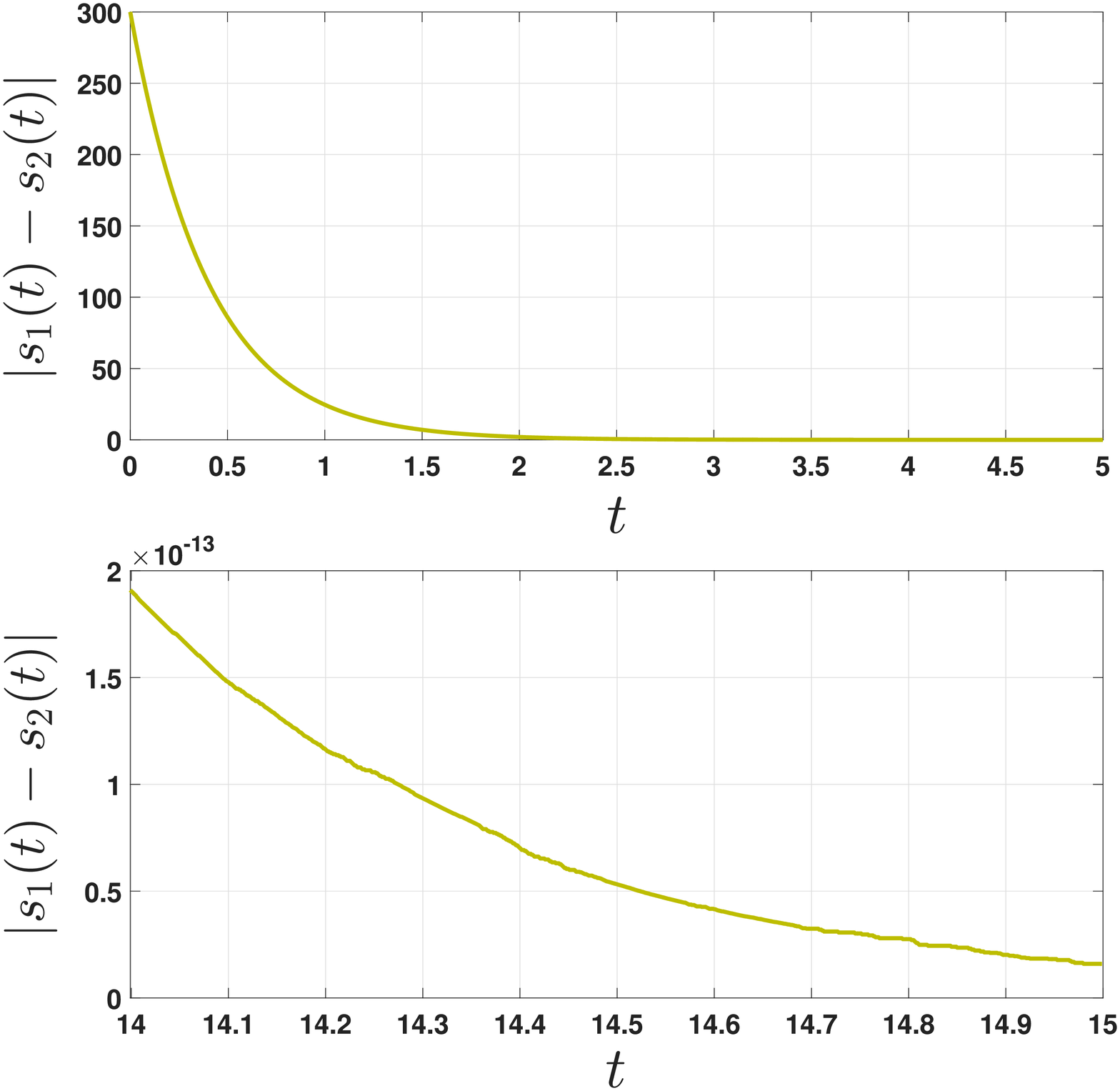}
  \caption{Synchronization error $|s_1(t) - s_2(t)|$. Responders are initialized in antiphase, i.e., $s_1(0) = - s_2(0)$.  }\label{Fig5}
\end{figure}

\begin{figure}[t]
  \centering
  \includegraphics[scale=.275]{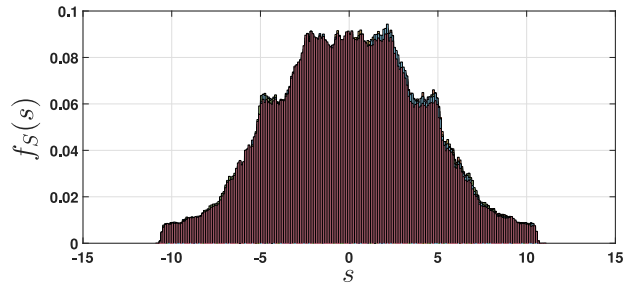}
  \caption{Empirical probability densities of samples, $s(t_k)$, from the synchronous solution $s_1(t) = s_1(t) = s(t)$, for twenty different, randomly selected, initial conditions.}\label{Fig6}
\end{figure}

\begin{figure}[t]
  \centering
  \includegraphics[scale=.275]{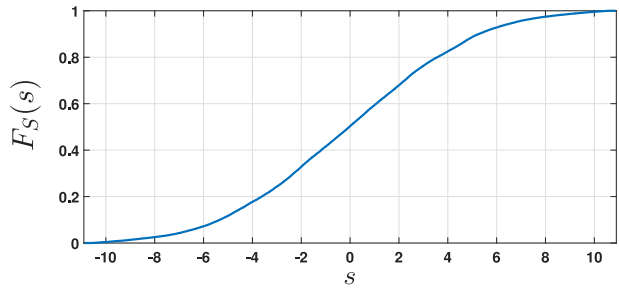}
  \caption{Empirical CDF corresponding to $f_S(s)$.}\label{Fig7}
\end{figure}

\begin{figure}[th]
  \centering
  \includegraphics[scale=.275]{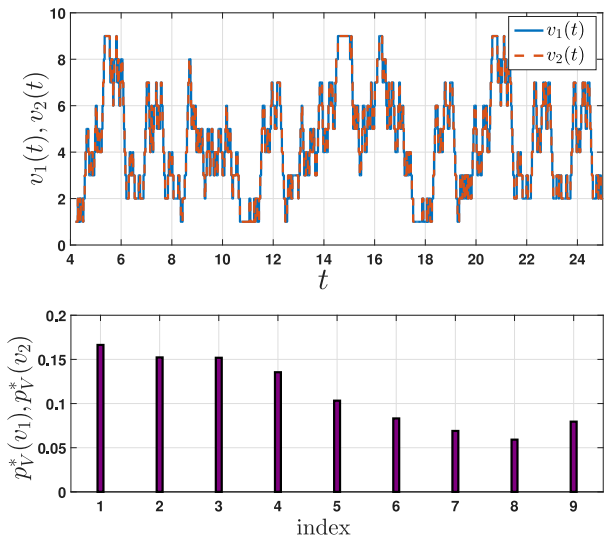}
  \caption{Top: realizations of $p^*_V(v)$ generated by the piecewise function \eqref{piecewise} at both sides of the channel, $v_1(t)$ at the trusted server and $v_2(t)$ at the remote station. Bottom: corresponding probability mass functions.}\label{Fig8}
\end{figure}

We next present an evaluation of our algorithms on real data. We use the \emph{adult-dataset}, available from the UCI Machine Learning Repository \cite{Dua:2019}, which contains census data. Each attribute within the dataset has $3.9 \times 10^4$ entries. We use three of these attributes: race, sex, and income, which take values on finite discrete sets. We let \emph{race} and \emph{sex} be the private information, $X$, and use \emph{income} as the information requested by the query, $Y$. The probability mass functions of $X$ and $Y$, and part of the one of $(X,Y)$ are given in Table \ref{table1}.\linebreak In Figure \ref{Fig3}, we depict $p_X(x)$, $p_Y(y)$, and $p_{X,Y}(x,y)$ with mass points indexed in the order given in Table \ref{table1}.\linebreak We first compute the optimal distribution $p_V^*(v)$ of the distorting additive noise $V$. We solve the convex program \eqref{eq:convex_optimization4} in Theorem \ref{th2}. The optimal distribution is depicted in Figure \ref{Fig_opt} and the corresponding numerical values are given in Table \ref{table2}. This $p_V^*(v)$ leads to $I[X;Y+V] = 0.0024$ while the mutual information without distortion is $I[X;Y] = 0.0251$, i.e., according to our metric, by optimally distorting the query, we leak about ten times less information. To generate realization from this distribution at both sides of the channel, we use trajectories of two chaotic responders as introduced in Section \ref{sync_problem}. We use the synthesis procedure in Section \ref{synth} to select suitable driver and responders. As driver \eqref{3}, we use the Lorenz system:
\begin{equation}\label{lorenz}
\left\{\begin{split}
\dot{\xi_1}(t) &= 10(\xi_2(t) - \xi_1(t)) ,\\
\dot{\xi_2}(t) &= 28\xi_1(t) - \xi_2(t) - \xi_1(t)\xi_3(t) ,\\
\dot{\xi_3}(t) &= -\tfrac{8}{3}\xi_3(t) + \xi_1(t)\xi_2(t),\\
u(t) &= \xi_1(t),
\end{split}\right.
\end{equation}
with states $\xi_1,\xi_2,\xi_3 \in \Real$ and driving signal $u \in \Real$. The Lorenz system produces bounded trajectories \cite{Pogr1}, and is known to be chaotic and quasi-hyperbolic \cite{Kapitaniak}. For the responders $\eqref{2}$ and $\eqref{4}$, we let $r(\zeta,u) = A\zeta + \psi(u)$, with $A=\text{diag}[-1,-2.5]$ and $\psi(u) = (-5u^2,50\sin(u))^\top$. Because $A$ is diagonal and has negative eigenvalues, responders satisfy the conditions of Corollary \ref{cor1} with $P = I_2$; hence, they are convergent systems and thus exponentially synchronize when driven by the same input $u(t)$. Moreover, since responders are linear in $\zeta$ and $A$ is Hurwitz, systems can be proved to be ISS with input $\psi(u)$ \cite{SONTAG}. Because $u$ is bounded and $\psi(u)$ is continuous, by the extreme value theorem, $\psi(u)$ is bounded, which, together with ISS, imply boundedness of the responders' trajectories \cite{SONTAG}. We let the outputs of the responders be $s_1(t) = \zeta_1^2$ and $s_2(t) = \zeta_2^2$ (their second state). In Figure \ref{Fig4}, we show traces of the chaotic driver and responders trajectories obtained by computer simulations (using Matlab from Mathworks), and in Figure \ref{Fig5}, we plot the synchronization error between the outputs of the responders. We initialized the responders in antiphase $\zeta_1(0) = - \zeta_2(0) = (150,150)^\top$, and far from the limit trajectory. Note, in Figure \ref{Fig5}, that responders synchronize exponentially and are practically synchronized for $t \geq 5$. Moreover, after $t \geq 14$, the synchronization error is within Matlab's precision ($10^{-12}$). Because the Lorenz system is quasi-hyperbolic, samples from the driving signal $u(t)$ follow a stationary distribution that is independent of the initial conditions of the driver, see Section \ref{chaos}. Then, according to the synthesis procedure in Section \ref{synth}, we next verify, using Monte Carlo simulations, that samples $s_k = s(t_k)$ (see Section \ref{numbers}), from the synchronous trajectory, $s_1(t) = s_1(t) = s(t)$, are also stationary. To do so, we compute the probability density function $f_S(s)$, $s_k \sim f_S(s)$, for different initial conditions and verify that all of them lead to the same density. In Figure \ref{Fig6}, we depict probability densities of $s_k$ for twenty different initial conditions, sampling instants $t_k = \Delta k$, $\Delta = 0.001$, and $t \in [0,4000]$. Note that they all lead to the same density $f_S(s)$. The support (obtained numerically) of $f_S(s)$ is given $\mathcal{S} = [-10.8585,10.8683]$. Finally, we use the piecewise function \eqref{piecewise} to generate realizations from $p_V^*(v)$ using samples, $s_k$, from the synchronous trajectory. Following the algorithm given in Section \ref{numbers}, we have to divide the support $\mathcal{S}$ of $f_S(s)$ into a set of partitions $C = \{c^1,\ldots,c^M \}$, such that the probability that $s_k$ lies in the cells equals the optimal probability distribution $p_V^*(v)$. This can be done using the empirical Cumulative Distribution Function (CDF), $F_S(s)$, corresponding to $f_S(s)$. We depict this CDF in Figure \ref{Fig7}. Then, we simply select the cells $C$ such that $p_V^*(y_i) = \text{Pr}[V = y_i] = \text{Pr}[c^i \leq S \leq c^{i+1}] = F_S(c^{i+1})-F_S(c^{i})$ for all $i \in \{1,\ldots,M-1\}$, $M = 9$ (the cardinality of the alphabet of $Y$). For this CDF and $p_V^*(v)$ in Table \ref{table2}, we obtain the following cells:
\begin{align}\label{cells}
C = \big\{ &[-\infty,-4.1739), [-4.1739,-2.0965), [-2.0965,-0.3658), [-0.3658,1.1408), [1.1408,2.3321) \\ &[2.3321,3.4341), [3.4341,4.5985), [4.5985,5.7743), [5.7743,\infty] \big\}.\notag
\end{align}
In Figure \ref{Fig8}, we show realizations generated by the piecewise function \eqref{piecewise} at both sides of the channel, and the corresponding probability mass functions. To generate this realizations, at the trusted server, we use samples from $s_1(t)$ and, at the remote station, we sample $s_2(t)$. Note that, as expected, all samples are perfectly synchronized and their probability mass functions are equal to $p_V^*(v)$ in Figure \ref{Fig_opt}.

\begingroup
\setlength{\tabcolsep}{1pt}
\begin{table}[h]
\centering
\begin{tabular}{|c|c|c|c|c|c|c|c|c|c|c|}
\hline
\addlinespace[.5ex]
$X$      & $\begin{bmatrix} 0 \\ 0 \end{bmatrix}$ & $\begin{bmatrix} 0 \\ 1 \end{bmatrix}$ & $\begin{bmatrix} 0 \\ 2 \end{bmatrix}$ & $\begin{bmatrix} 0 \\ 3 \end{bmatrix}$ & $\begin{bmatrix} 0 \\ 4 \end{bmatrix}$ & $\begin{bmatrix} 1 \\ 0 \end{bmatrix}$ & $\begin{bmatrix} 1 \\ 1 \end{bmatrix}$ & $\begin{bmatrix} 1 \\ 2 \end{bmatrix}$ & $\begin{bmatrix} 1 \\ 3 \end{bmatrix}$ & $\begin{bmatrix} 1 \\ 4 \end{bmatrix}$ \\ \addlinespace[.5ex] \hline \addlinespace[.5ex]
$p_X(x)$ & 0.5888 & 0.0200 & 0.0056 & 0.0560 & 0.0038 & 0.2616 & 0.0110 & 0.0042 & 0.0468 & 0.0022 \\ \addlinespace[.5ex] \hline
\end{tabular}\\[3mm]
\begin{tabular}{|c|c|c|c|c|c|c|c|c|c|}
\hline
\addlinespace[.5ex]
$Y$ & 1 & 2 & 3 & 4 & 5 & 6 & 7 & 8 & 9 \\ \addlinespace[.5ex] \hline \addlinespace[.5ex]
$p_Y(y)$ & 0.6870 & 0.0766 & 0.0364 & 0.0292 & 0.0658 & 0.0386 & 0.0002 & 0.0001 & 0.0662 \\ \addlinespace[.5ex] \hline
\end{tabular}\\[3mm]
\begin{tabular}{|c|c|c|c|c|c|c|c|c|c|c|c|}
\hline
\addlinespace[.5ex]
$(X,Y)$      & $\begin{bmatrix} 0 \\ 0 \\ 1 \end{bmatrix}$ & $\begin{bmatrix} 0 \\ 1 \\ 1  \end{bmatrix}$ & $\begin{bmatrix} 0 \\ 2 \\ 1 \end{bmatrix}$ & $\begin{bmatrix} 0 \\ 3 \\ 1 \end{bmatrix}$ & $\begin{bmatrix} 0 \\ 4 \\ 1 \end{bmatrix}$ &  $\cdots$ & $\begin{bmatrix} 1 \\ 0 \\ 9 \end{bmatrix}$ & $\begin{bmatrix} 1 \\ 1 \\ 9 \end{bmatrix}$ & $\begin{bmatrix} 1 \\ 2 \\ 9 \end{bmatrix}$ & $\begin{bmatrix} 1 \\ 3 \\ 9 \end{bmatrix}$ & $\begin{bmatrix} 1 \\ 4 \\ 9 \end{bmatrix}$ \\ \addlinespace[.5ex] \hline \addlinespace[.5ex]
$p_{X,Y}(x,y)$ & 0.3974 & 0.0130 & 0.0044 & 0.0388 & 0.0032 & \hspace{.5mm}$\cdots$ & 0.0222 & 0.0014 & 0.0008 & 0.0046 & 0.0004 \\ \addlinespace[.5ex] \hline
\end{tabular}
\caption{Probability mass functions of $X$ and $Y$, and part of the one of $(X,Y)$.}
\label{table1}
\end{table}
\endgroup

\section{Conclusions}

Using an information-theoretic privacy metric (mutual information), we have provided a general privacy framework based on additive distorting random vectors and exponential synchronization of chaotic systems. The synthesis of the optimal probability distribution, $p^*_V(v)$, of the additive distorting vector $V$ has been posed as a convex program in $p_V(v)$. We have provided an algorithm for generating pseudorandom realizations from this distribution using trajectories of chaotic oscillators. To generate equal realizations at both sides of the channel, we have induced exponential synchronization on two chaotic oscillators (one at each side of the channel), and use their trajectories and the proposed algorithm to generate realizations. However, exponential synchronization implies that, in finite time, there is always a small error between trajectories (and thus also between realizations). We have derived an upper bound on the worst-case distortion induced by finite-time synchronization errors and showed that this distortion disappears exponentially fast. Using off-the-shelf results in the literature, we have provided general guidelines for selecting the dynamics of the responders and driver so that our algorithm for generating synchronized realizations from $p^*_V(v)$ is guaranteed to work. We have presented simulation results to illustrate our results.

\begingroup
\setlength{\tabcolsep}{1pt}
\begin{table}[H]
\centering
\begin{tabular}{|c|c|c|c|c|c|c|c|c|c|}
\hline
\addlinespace[.5ex]
$V$ & 1 & 2 & 3 & 4 & 5 & 6 & 7 & 8 & 9 \\ \addlinespace[.5ex] \hline \addlinespace[.5ex]
$p^*_V(v)$ & 0.1664 & 0.1522 & 0.1518 & 0.1355 & 0.1033 & 0.0832 & 0.0690 & 0.0591 & 0.0795 \\ \addlinespace[.5ex] \hline
\end{tabular}\\[3mm]
\caption{Optimal distribution $p_V^*(v)$ of the distorting additive random variable $V$.}
\label{table2}
\normalsize
\end{table}
\endgroup

\bibliographystyle{IEEEtran}
\bibliography{ifacconf32}

\end{document}